\documentclass[conference]{IEEEtran}
\IEEEoverridecommandlockouts
\usepackage{cite}
\usepackage{amsmath,amssymb,amsfonts}
\usepackage{algorithmic}
\usepackage{subcaption}
\usepackage{graphicx}
\usepackage{textcomp}
\usepackage{multirow}
\usepackage{pifont}
\usepackage{xcolor}
\usepackage[hidelinks]{hyperref}
\usepackage{makecell}
\usepackage{booktabs}
\usepackage{array}
\usepackage{url}
\def\BibTeX{{\rm B\kern-.05em{\sc i\kern-.025em b}\kern-.08em
    T\kern-.1667em\lower.7ex\hbox{E}\kern-.125emX}}
\urldef{\PowerAgentBenchURL}\url{https://github.com/Power-Agent/PowerAgentBench}

\begin{document}

\title{PowerAgentBench-Dyn: A Benchmark for Agentic AI in Power System Dynamic Studies\\
\thanks{Q. Zhang and L. Xie are with the School of
Engineering and Applied Sciences, Harvard University, Allston, MA, USA. A. Pomarico and A. Berizzi are with Department of Energy, Politecnico di Milano, Milan, Italy. C. Mylonas and M. Foti are with Energy Digitalization Group, UBITECH, Athens, Greece. Corresponding authors with equal contribution: Qian Zhang and Andrea Pomarico (qianzhang@g.harvard.edu, andrea.pomarico@polimi.it). 

The code repository of this paper is maintained by PowerAgent Community at: \PowerAgentBenchURL.}}

\author{\IEEEauthorblockN{Qian Zhang$^{*}$, Andrea Pomarico$^{*}$, Costas Mylonas, Magda Foti, Alberto Berizzi, Le Xie}}

\maketitle

\begin{abstract}
Large Language Model (LLM)-based agents are increasingly being used to automate multi-step engineering workflows by interacting with software tools, interpreting intermediate results, and autonomously planning subsequent actions. Power system dynamic studies represent a particularly promising yet largely unexplored application domain for these agents. Unlike static computational tasks, dynamic studies often require more time on model parameter calibration, engineering judgment, and decision making under constrained action spaces.
This paper introduces \textit{PowerAgentBench-Dyn}, a benchmark designed to evaluate Agentic AI systems on power system dynamic-analysis tasks. The benchmark targets problems that cannot be reduced to a single optimization or coding task, but instead require a type of reasoning, tool usage, and iterative experimentation routinely performed by experienced power system engineers.
The proposed framework includes two \emph{initial} benchmark tasks. The first, the \textit{Dynamic Model Quality Review Benchmark}, evaluates agents' ability to validate and diagnose dynamic models based on model-quality compliance criteria specified by system operators. The second, the \textit{Dynamic Security Risk Screening Benchmark}, assesses agents' capability to leverage semantic memory and a limited simulation budget to identify, rank, and analyze the most critical short-circuit contingencies from an unseen fault dataset, as well as propose and evaluate possible mitigation measures.
For each task, we define the simulation environment, observation and action spaces, and evaluation metrics. The benchmark is reproducible in a metric-based sense: released cases and simulator settings define a deterministic evaluator, while stochastic agent behavior is assessed over repeated runs using success rates and other metrics. The benchmark supports the development of future Agentic AI for power system operation and planning.

\end{abstract}


\section{Introduction}

Power system dynamic studies are becoming increasingly important as inverter-based resources (IBRs), storage, data centers, and weather-dependent operating conditions reshape the stability characteristics of modern grids. Recent IBR disturbances have highlighted the need for better dynamic model quality, model validation, and disturbance replication workflows \cite{nerc2022odessa}. Meanwhile, the advances in agentic AI have demonstrated that language models equipped with reasoning and tool-use capabilities can move beyond single-turn question answering, enabling code execution, feedback-driven action revision, and multi-step task solving \cite{react,agentbench,toolllm,swebench}. This capability is especially relevant for some power engineering workflows where the bottleneck is not a single formula, but a sequence of decisions: preparing cases, selecting disturbances, running simulations, diagnosing failed runs, interpreting plots, changing parameters within allowed ranges, and producing an auditable report.

The application of agentic AI in power system steady-state analysis has been summarized in our complementary benchmark \emph{PowerAgentBench-SS} \cite{mylonas2026poweragentbenchss}. The exploration of applying Agentic AI in power system dynamic study is still at early stage. Existing AI applications in power system dynamics have largely focused on supervised learning for transient stability assessment, contingency classification, or dynamic security assessment \cite{10858603,mehrzad2023review}. These approaches are valuable when a fixed training distribution and well-defined labels are available. However, many daily engineering tasks are open-ended and operationally constrained. For example, a dynamic model reviewer may need to determine why a plant model does not match PMU data, choose which model quality test to run next, and propose a parameter change without modifying locked files \cite{sun2024inverter}. A planning engineer may need to identify the ten most concerning short circuits among a large pool of possible fault locations and clearing times, even though only a subset can be simulated within the available time \cite{ni2003online}.

This paper proposes \textit{PowerAgentBench-Dyn}, a benchmark suite for evaluating agentic AI on power system dynamic studies. Our design principle is to focus on tasks that are hard to solve by deterministic scripts or single-shot optimization because they require multi-round interaction with simulation tools and engineering judgement under a user-defined action space. The contributions of this paper are threefold:
\begin{itemize}
    \item We formulate a general agentic benchmark framework for dynamic studies, including observations, action spaces, simulator feedback, constraints, and evaluation metrics.
    \item We design a Dynamic Model Quality Review Benchmark inspired by industrial model quality testing workflows \cite{cheng2025dmview}, including the model quality testing and iterative mitigation process.
    \item We design a Dynamic Security Risk Screening Benchmark in which an agent uses semantic memory and a limited simulation budget to rank high-risk short-circuit contingencies from an unseen fault dataset, with an extension toward autonomous mitigation selection.
\end{itemize}

\section{Overview of Power System Dynamic Studies}

\subsection{Category and Simulation Software}

Power system dynamic studies evaluate the time-domain response of a network after disturbances such as faults, line trips, generator trips, frequency events, or control actions. Typical study categories include transient stability, voltage stability, frequency response, oscillation damping, dynamic model validation, Model Quality Testing (MQT), ride-through testing, and Electromagnetic Transient (EMT) studies. These studies are usually performed with commercial or open-source simulation environments such as PSS/E, DIgSILENT PowerFactory, PSCAD, TSAT, PSLF, ANDES, and other domain-specific tools. Positive-sequence root-mean-square (RMS) simulation is commonly used for large-scale planning and operations studies, while EMT simulation is often used for detailed IBR control and protection interactions.

Dynamic studies are fundamentally different from steady-state power flow or contingency analysis. A dynamic case includes differential-algebraic models, protection logic, controllers, event sequences, integration settings, and output channels. A failed simulation may indicate a true physical instability, a numerical issue, an incorrect model parameter, or an initialization problem. Therefore, most engineer's dynamic study workflow is iterative: inspect the case, run a baseline, test disturbances, review waveforms, distinguish physical and modeling causes, modify only allowed inputs, and document the result.

\subsection{The Value of Engineering Experience in Dynamic Studies}

Experienced engineers contribute knowledge that is difficult to encode as a fixed deterministic program. For model quality review, they know which tests should be run first, what a flat-start drift implies, whether a voltage recovery shape is plausible, and which parameters are likely related to active power recovery, reactive current injection, frequency response, or low-voltage power logic. For security screening, they know that the most severe contingencies are not always the longest faults or the electrically closest locations; severity depends on operating condition, protection clearing, grid strength, synchronizing torque, damping, and the location of critical machines or IBR plants.

This motivates an agentic benchmark rather than a conventional prediction benchmark. The agent is not only asked to classify a static feature vector. Instead, it receives a task objective, a simulator interface, a set of user constraints, and an action budget. It must decide which simulation or diagnostic action to perform next, interpret the result, and stop with a final engineering judgement.

\subsection{Framework for Agentic AI Benchmark}

We represent each benchmark task as an interactive environment
\begin{equation}
\mathcal{T} = (\mathcal{O}, \mathcal{A}, \mathcal{E}, \mathcal{C}, \mathcal{G}, \mathcal{M}),
\end{equation}
where $\mathcal{O}$ is the observation space, $\mathcal{A}$ is the constrained action space, $\mathcal{E}$ is the simulation environment, $\mathcal{C}$ is the set of user and engineering constraints, $\mathcal{G}$ is the hidden ground truth or evaluator, and $\mathcal{M}$ is optional semantic memory. At step $t$, the agent receives observation $o_t$, selects action $a_t \in \mathcal{A}$, obtains simulator or tool feedback $o_{t+1}$, and continues until it returns a final answer or reaches the action/time budget.

The action space is intentionally constrained. In real dynamic studies, an engineer cannot arbitrarily modify confidential models or ignore grid operator criteria. Each benchmark instance therefore includes a machine-readable constraint file specifying allowed parameters, allowed event modifications, maximum simulations, maximum wall-clock time, locked files, and reporting requirements.

\begin{figure}[h]
\centering
\includegraphics[width=\columnwidth]{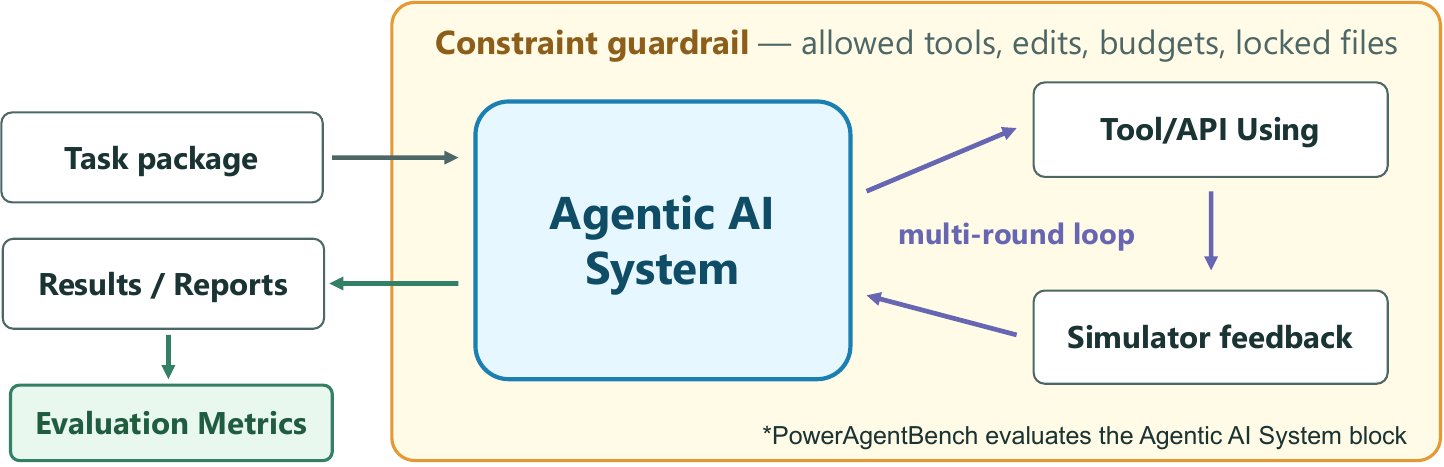}
\caption{PowerAgentBench framework overview}
\label{fig:framework_overview}
\end{figure}

As illustrated in Fig. \ref{fig:framework_overview}, the benchmark framework is designed based on 4 principles.
\begin{enumerate}
    \item \textit{Tool-grounded evaluation}: agents must interact with simulators and analysis tools provided by users.
    \item \textit{Multi-round decision making}: the task should require more than single round response or deterministic script.
    \item \textit{Engineering constraint guardrail}: actions must remain within allowed action space, budgets, and files.
    \item \textit{Outcome-oriented metrics}: success is measured by reliability-relevant outcomes such as diagnosis accuracy, waveform fit, ranking quality, mitigation success, and wall-clock solving time.
\end{enumerate}

\section{Dynamic Model Quality Review Benchmark}

\subsection{Task Description}

Dynamic model quality review is a central workflow for interconnection and reliability studies. A submitted renewable resource or large load dynamic model may initialize in a planning case, but still fail model quality tests because of unstable controller gains, incorrect ride-through behavior, weak-grid sensitivity, or poor active/reactive power recovery. The proposed Dynamic Model Quality Review Benchmark (DMQ Benchmark) evaluates whether an AI agent can use a dynamic-model review tool, diagnose the failed responses, and make only user-authorized model-file edits.

\begin{table}[t]
\caption{Configuration of the DMQ benchmark.}
\label{tab:task_config_dmqr}
\centering
\scriptsize
\setlength{\tabcolsep}{3pt}
\renewcommand{\arraystretch}{1.1}
\begin{tabular}{@{}>{\raggedright\arraybackslash}p{0.22\columnwidth}
                 >{\raggedright\arraybackslash}p{0.72\columnwidth}@{}}
\toprule
\textbf{Item} & \textbf{Description} \\
\midrule
Tools & DMView, PSS/E dynamic simulation. \\
Inputs & \texttt{Solar.dyr}, \texttt{Solar.sav}, DMView test configuration. \\
Allowed actions & Modify only the four REECAU1 gains, within a budget of five iterations. \\
Metrics & Full-suite pass rate, waveform fit, LVRT recovery, wall-clock time, constraint compliance. \\
\bottomrule
\end{tabular}
\end{table}

As summraized in TABLE \ref{tab:task_config_dmqr}, the DMView functions are exposed to the agent as tools, which is an open-source PSS/E dynamic model review platform that supports flat-start tests, voltage and frequency step tests, high- and low-voltage ride-through tests, and model validation workflows \cite{cheng2025dmview,cheng2018dmview}. The agent first connects the user provided model to the DMView test system and runs the required test suite. If one or more tests fail, the agent reads the DMView pass/fail report, simulation logs, and waveform-derived diagnostics, then proposes a constrained update to the PSS/E \texttt{.dyr} model file. The update is applied only if it is inside the user-provided action space. The agent then reruns the test suite and iterates until all tests pass or the maximum number of iterations is reached.

As an initial task, the action space is deliberately narrow. For the case study below, the agent may modify only 4 REECAU1 controller gains in the data row following the \texttt{REECAU1} model declaration: columns 24--27, interpreted as $K_{qp}$, $K_{qi}$, $K_{vp}$, and $K_{vi}$. The agent is not allowed to change model structure, bus numbers, machine identifiers, current limits, voltage thresholds, protection flags, plant-controller parameters, or any other \texttt{.dyr} fields. This design turns the task into a constrained engineering workflow rather than an unconstrained parameter search.

\subsection{Evaluation Metrics}

The DMQ Benchmark evaluates both the final model and the process used to obtain it. The primary metric is full-suite success: whether the final model passes all required DMView tests within the iteration budget. Secondary metrics include the number of DMView simulations, pass/fail status for each individual test, LVRT momentary-cessation duration, active/reactive power drift in flat start, peak-to-peak oscillation magnitude, weak-grid SCR performance, and quality of the final engineering report. In practice, users can select any individual metric or combination of metrics according to their preferences.

Constraint compliance is evaluated separately from test success. An agent receives credit only if it modifies whitelisted parameters, preserves locked fields, records each parameter change, and provides an evidence-based reason for each update. This is important because a model can appear improved after an unauthorized edit, but such a solution would not be acceptable in a real interconnection or model-validation workflow. 

\subsection{Case Study: WECC Solar Model}

The case study uses a modified WECC solar photovoltaic dynamic model with an overly aggressive REECAU1 gain set:
\[
(K_{qp},K_{qi},K_{vp},K_{vi})=(10,50,10,50).
\]
The input package contains a PSS/E power-flow case, the modified \texttt{Solar.dyr} file, the DMView test configuration, and an instruction file defining the allowed action space and a \emph{5-iteration budget}. The test suite comprises 8 model-quality tests which are provided by DMView package. 

At the baseline, the model fails every test in the suite. The observed symptoms include flat-start drift of approximately 1.80 pu in active power and 0.82 pu in reactive power, reactive-power oscillations of approximately 1.3--1.6 pu peak-to-peak in the voltage tests, active-power collapse in the frequency tests, and a low-voltage ride-through (LVRT) momentary-cessation event lasting roughly 19 s. These failures are consistent with excessive reactive and voltage control gains driving an unstable Q/V control loop.

We mainly compare the performance from 3 different Anthropic models, where the same input package using an automated harness that exposes the DMView suite as tools, enforces the 4-parameter action space, and records every model edit. Each model was run for ten independent sessions\footnote{The full per-run gain trajectories are available in the code repository.} As summarized in Table \ref{tab:dmq_case_results}, Opus 4.8 and Sonnet 4.6 both repair the model in every session, typically with a single parameter update that reduces all four gains by roughly an order of magnitude. Both models reliably damp the Q/V oscillations and restore prompt LVRT recovery on the first attempt. Haiku 4.5 succeeds in 9 of 10 sessions but is the most discriminating tier: it tends to approach a passing set through several smaller reductions (median of two updates, and occasionally the full five-iteration budget), and in one session it fails to converge within the budget, repeatedly retuning the integral gains without removing the residual weak-grid oscillation. This contrast is useful because it shows that the benchmark separates models not on whether a fix exists, but on how efficiently and reliably they localize the allowed parameters and converge within a constrained action space.

\begin{table}[h]
\caption{DMQ case-study results on the modified WECC solar model (over 10 independent runs per model).}
\label{tab:dmq_case_results}
\centering
\scriptsize
\setlength{\tabcolsep}{2pt}
\begin{tabular}{@{}
>{\centering\arraybackslash}p{0.18\columnwidth}
>{\centering\arraybackslash}p{0.20\columnwidth}
>{\centering\arraybackslash}p{0.18\columnwidth}
>{\centering\arraybackslash}p{0.34\columnwidth}
@{}}
\toprule
Agent & Success rate & Median iter. & Median final gains \\
\midrule
Opus 4.8 & 10/10 & 1 & $(1,5,1,5)$ \\
Sonnet 4.6 & 10/10 & 1 & $(0.5,2,0.5,3.5)$ \\
Haiku 4.5 & 9/10 & 2 & $(1.2,2,1.5,4)$ \\
\bottomrule
\end{tabular}
\end{table}

The reported results should be viewed as the benchmark performance based on a single LLM agent, rather than as the only possible agentic system design. A user could build a more advanced agentic system by adding engineering skills, retry logic for failed tool calls, or ensemble voting across repeated LLM calls. PowerAgentBench-Dyn keeps the case, action constraints, and scoring fixed so that such agent designs can be compared under the same evaluation protocol.

\section{Dynamic Security Risk Screening Benchmark}

\subsection{Task Description}

Dynamic security risk screening (DSR) aims to identify the disturbances that are most likely to compromise power system stability, including transient instability, poorly damped oscillations, voltage recovery violations, frequency excursions, and IBR ride-through issues. In principle, this requires simulating all possible combinations of operating conditions, fault locations, fault types, protection clearing times, and post-fault topologies. In practice, however, the number of scenarios becomes prohibitively large, while dynamic simulations are computationally expensive and operators often require timely risk assessments.

\begin{table}[t]
\caption{Configuration of the Dynamic Security Risk Screening benchmark.}
\label{tab:task_config_dsrs}
\centering
\scriptsize
\setlength{\tabcolsep}{3pt}
\renewcommand{\arraystretch}{1.1}
\begin{tabular}{@{}>{\raggedright\arraybackslash}p{0.22\columnwidth}
                 >{\raggedright\arraybackslash}p{0.72\columnwidth}@{}}
\toprule
\textbf{Item} & \textbf{Description} \\
\midrule
Tools & PowerMCP, DIgSILENT PowerFactory. \\
Inputs & Semantic memory, unseen candidate faults, simulation KPIs, mitigation options. \\
Allowed actions & Query the semantic memory, a limited number of simulations, evaluate SC placements within the mitigation budget. \\
Metrics & Top-$k$ ranking quality, severity score, mitigation success rate, wall-clock time, constraint compliance. \\
\bottomrule
\end{tabular}
\end{table}

As summarized in TABLE \ref{tab:task_config_dsrs}, the DSR benchmark evaluates a closed-loop screening workflow: the agent queries semantic memory, selects contingencies to simulate, computes dynamic-security KPIs, and tests mitigation options under a fixed action budget. The agent operates through the open-source Model Context Protocol (MCP) framework implemented within the PowerAgent platform \cite{zhang2025poweragent}, which connects the LLM to DIgSILENT PowerFactory. This controlled setting reproduces the experience-driven workflow used by TSOs and RTOs while keeping the candidate set, memory content, simulation budget, and allowed mitigation actions fixed for evaluation.

\subsection{Evaluation Metrics}
For this task, reproducibility does not require identical natural-language traces across runs. Instead, distance to the deterministic ground truth is measured by top-$k$ overlap and ranking quality for critical-contingency selection, absolute errors in severity score and voltage extrema, mitigation success, wall-clock time, and constraint compliance. A run is counted as workflow-success only if the agent completes the ranking, simulation, and mitigation stages without violating the action budget or allowed mitigation set. To quantitatively assess each simulation outcome and support the ranking process, the following Key Performance Indicators (KPIs) are adopted:

\begin{itemize}
    \item \textbf{Bus voltage extrema.} The max and min per-unit bus voltages observed during the post-fault recovery window, denoted as $V_{\max}^{\mathrm{aft}}$ and $V_{\min}^{\mathrm{aft}}$, together with the corresponding buses. These indicators capture localized over- and under-voltage conditions following the disturbance.



    \item \textbf{Stability flag.} A binary indicator classifying the post-fault system response as stable or unstable.

    \item \textbf{Severity score.} A scalar indicator $S \in [0,1]$, where higher values correspond to more severe contingencies. The score is computed using an \emph{Integral of Violation} (IoV) approach based on post-fault voltage deviations. For each bus, the normalized under- and over-voltage violations are integrated over time and averaged across all $X$ buses. Denoting by $\mathrm{IoV}_{lo}(x)$ and $\mathrm{IoV}_{hi}(x)$ the normalized under- and over-voltage integrals for bus $x$, the partial scores are defined as
    \begin{equation}
        s_{lo} = \mathrm{clip}_{[0,1]}\!\left(\frac{1}{X}\sum_{x=1}^{X}\mathrm{IoV}_{lo}(x)\right),
    \end{equation}

    \begin{equation}
        s_{hi} = \mathrm{clip}_{[0,1]}\!\left(\frac{1}{X}\sum_{x=1}^{X}\mathrm{IoV}_{hi}(x)\right),
    \end{equation}

    where the per-bus violations are normalized between activation and saturation thresholds ($0.95 \rightarrow 0.90$~pu for under-voltage and $1.05 \rightarrow 1.10$~pu for over-voltage). The final severity score is computed as
    \begin{equation}
        S =
        \begin{cases}
            1, & \text{if system is unstable,}\\[4pt]
            w_{lo}\, s_{lo} + w_{hi}\, s_{hi}, & \text{otherwise,}
        \end{cases}
    \end{equation}
    where equal weights are adopted, i.e., $w_{lo}=w_{hi}=0.5$. The instability override ensures that contingencies leading to instability always receive the maximum severity score.
\end{itemize}

\subsection{Case Study}
The case study evaluates Claude Opus 4.7, Sonnet 4.6, and Haiku 4.5 by Anthropic, Gemini 3.1 Pro by Google, GPT-5.5 by OpenAI, and the open-weights Qwen2.5-Coder-7B model developed by Alibaba Cloud. Each model is run in ten independent sessions using the same benchmark inputs, semantic memory, simulator settings, and action constraints. Because LLM policies are stochastic, we report workflow success rates and median KPI estimates rather than requiring identical action transcripts.\footnote{The corresponding box plots are available in the GitHub repository.}

The input prompt can be summarized in seven sequential steps:

\begin{itemize}
    \setlength{\itemsep}{0pt}
    \setlength{\parskip}{0pt}
    \setlength{\parsep}{0pt}
    \item \textbf{Step 1} Read a CSV file containing the ten contingencies and query the semantic memory to verify whether the corresponding simulations are already available.
    
    \item\textbf{Step 2} If the simulations are not present in the memory, retrieve similar cases based on network topology, initial operating conditions, fault location, and fault duration.
    
    \item\textbf{Step 3} Use the retrieved knowledge to estimate the KPIs for each contingency.
    
    \item \textbf{Step 4} Rank the contingencies according to the KPIs.
    
    \item\textbf{Step 5} Select the three most critical contingencies and simulate them in DIgSILENT PowerFactory, storing the results and computing the corresponding KPIs.
    
    \item \textbf{Step 6} Analyze the simulation outcomes and compare them with the estimated values to validate or refine the initial reasoning.
    
    \item \textbf{Step 7}  Evaluate possible countermeasures for the simulated contingencies, summarize their effects, and identify the solution with the greatest impact.
\end{itemize}
The considered contingencies are listed below:

\begin{itemize}
    \setlength{\itemsep}{0pt}
    \setlength{\parskip}{0pt}
    \setlength{\parsep}{0pt}
    \item Three-phase short circuit at Bus 14 lasting 160 ms
    \item Three-phase short circuit at Bus 18 lasting 100 ms
    \item Three-phase short circuit at Bus 05 lasting 180 ms
    \item Three-phase short circuit at Bus 19 lasting 20 ms
    \item Three-phase short circuit at Bus 15 lasting 100 ms
    \item Three-phase short circuit at Line 05--06 lasting 200 ms
    \item Three-phase short circuit at Line 02--25 lasting 160 ms
    \item Three-phase short circuit at Line 06--11 lasting 60 ms
    \item Three-phase short circuit at Line 04--05 lasting 20 ms
    \item Three-phase short circuit at Line 08--09 lasting 200 ms
\end{itemize}

All ten contingencies are simulated offline to define the deterministic ground truth. The three most critical contingencies are L02--25@160~ms, L05--06@200~ms, and B05@180~ms. A trial is marked successful if the agent completes the tool workflow, selects these critical cases for simulation, correctly flags the instability at L02--25@160~ms, and evaluates only the allowed SC mitigation options. The resulting success rates over ten independent trials are reported in Table~\ref{tab:success_rate}.
\begin{table}[t]
\caption{Workflow success rate achieved by each LLM.}
\label{tab:success_rate}
\centering
\resizebox{\columnwidth}{!}{%
\begin{tabular}{lcccccc}
\toprule
&
\textbf{Opus 4.7} &
\textbf{Sonnet 4.6} &
\textbf{Haiku 4.5} &
\textbf{Gemini 3.1 Pro} &
\textbf{GPT-5.5} &
\textbf{Qwen 7B} \\
\midrule
Success rate (\%) & 90 & 90 & 70 & 100 & 100 & 0 \\
\bottomrule
\end{tabular}}
\end{table}

Opus 4.7, Sonnet 4.6, Gemini 3.1 Pro, and GPT-5.5 reliably complete the workflow and identify the same critical contingencies; Haiku 4.5 completes 70\% of trials, while Qwen7B fails to complete the MCP-based pipeline.\footnote{Qwen7B is the smallest tested Qwen variant; larger models were not tested because the available workstation has only 4~GB of VRAM.} Failed Haiku and Qwen runs are mainly blocked by tool execution in DIgSILENT PowerFactory rather than by incorrect severity reasoning. As reported in TABLE \ref{tab:kpi_comparison}, completed proprietary-model runs correctly flag the L02--25@160~ms instability and estimate the remaining voltage-violation KPIs with small absolute errors.

The mitigation stage considers Synchronous Condenser (SC) installations at buses 04, 16, and 26, with a budget of two 250~MVA units. Because these device configurations are new planning actions, the agent does not query semantic memory and must evaluate the options through fresh simulations. As shown in TABLE \ref{tab:sc_comparison}, completed LLM runs consistently find that SC04 alone is ineffective and that any stabilizing solution must include SC26 for the L02--25@160~ms contingency.

Runtime is included as a process metric because a ranking that is accurate but too slow may be less useful for operational screening. Median completion times are 05:43 for GPT-5.5, 05:45 for Gemini 3.1 Pro, 07:05 for Opus 4.7, 07:22 for Sonnet 4.6, and 11:47 for Haiku 4.5, indicating that successful proprietary-model runs finish within minutes.

\begin{table}[t]
\caption{Ground truth and LLM estimation of the KPIs for the three most critical contingencies. (voltages in pu)}
\label{tab:kpi_comparison}
\centering
\resizebox{\columnwidth}{!}{%
\begin{tabular}{ll|ccc}
\toprule
\toprule
\textbf{Case} & \textbf{Method} &
\textbf{$S$} &
\textbf{$V_{\max}^{\mathrm{aft}}$@Bus} &
\textbf{$V_{\min}^{\mathrm{aft}}$@Bus} \\
\midrule
L02-25@160 & All LLMs & 1 & -- & -- \\
\midrule
\multirow{6}{*}{L05-06@200}
 & Ground truth    & 0.053 & 1.1623@29 & 0.6554@31 \\
 & Opus 4.7 & 0.043 & 1.1622@29 & 0.6968@31 \\
 & Sonnet 4.6 & 0.040 & 1.1621@29 & 0.6971@31 \\
 & Haiku 4.5  &0.034 & 1.1618@29 & 0.679@31 \\
 & Gemini 3.1 Pro  & 0.038 & 1.16@29   & 0.72@31 \\
 & GPT-5.5         & 0.045 & 1.1620@29 & 0.6950@31 \\
\midrule
\multirow{6}{*}{B05@180}
 & Ground truth    & 0.0342 & 1.1527@29 & 0.7723@31 \\
 & Opus 4.7 & 0.0356 & 1.1527@29 & 0.7658@31 \\
 & Sonnet 4.6 & 0.0372 & 1.153@29 & 0.772@31 \\
 & Haiku 4.5  & 0.031 & 1.1494@29 & 0.758@31 \\
 & Gemini 3.1 Pro  & 0.0350 & 1.15@29   & 0.77@31 \\
 & GPT-5.5         & 0.0356 & 1.1528@29 & 0.7659@31 \\
\bottomrule
\end{tabular}}
\par\smallskip
\end{table}

\begin{table}[t]
\caption{Dynamic security assessment of contingency L02-25@160 for different SC combinations.}
\label{tab:sc_comparison}
\centering
\scriptsize
\setlength{\tabcolsep}{2.5pt}
\renewcommand{\arraystretch}{0.95}
\begin{tabular}{ll|ccc}
\toprule
\toprule
\textbf{Case} & \textbf{Method} &
\textbf{$S$} &
\textbf{$V_{\max}^{\mathrm{aft}}$@Bus} &
\textbf{$V_{\min}^{\mathrm{aft}}$@Bus} \\
\midrule
\makecell[l]{SC04, SC16,\\ SC04+SC16} & All LLMs & 1 & -- & -- \\
\midrule
\multirow{5}{*}{SC26}
 & Opus 4.7   & 0.0839 & 1.2198@29 & 0.7934@31 \\
 & Sonnet 4.6 & 0.084 & 1.22@29 & 0.79@31 \\
 & Haiku 4.5  & 0.084 & 1.219@29 & 0.793@31 \\
 & Gemini 3.1 Pro    & 0.084  & 1.219@29  & 0.793@31 \\
 & GPT-5.5           & 0.0839 & 1.2198@29 & 0.7934@31 \\
\midrule
\multirow{5}{*}{SC04+SC26}
 & Opus 4.7   & 0.932  & 1.2274@29 & 0.7501@31 \\
 & Sonnet 4.6 & 0.93  & 1.227@29 & 0.75@31 \\
 & Haiku 4.5  & 0.932  & 1.23@29 & 0.75@31 \\
 & Gemini 3.1 Pro    & 0.932  & 1.227@29  & 0.750@31 \\
 & GPT-5.5           & 0.932  & 1.2274@29 & 0.7501@31 \\
\midrule
SC16+SC26 & All LLMs & 0.0830 & 1.2147@29 & 0.7709@31 \\
\bottomrule
\end{tabular}
\end{table}

These DSR results are benchmark outcomes, not an upper bound on agentic reproducibility. The benchmark fixes the faults, memory, simulator interface, actions, and metrics; users can still design more reproducible agents through deterministic planners, retrieval validators, tool-call checks, and repeated-run aggregation.

\section{Discussion}

\subsection{Why Agentic Evaluation is Needed}

The two proposed tasks are intentionally different from conventional supervised learning benchmarks. In a supervised transient stability classifier, the model receives a fixed feature vector and returns a label. In PowerAgentBench-Dyn, the agent receives an engineering objective and must decide how to use tools. This enables evaluation of capabilities that are critical for practical deployment: planning, tool selection, simulator error recovery, interpretation of plots, constraint following, and report generation.

This design also avoids over-emphasizing token cost or single-shot accuracy. For dynamic studies, a slightly slower agent that produces a correct, auditable, and constraint-compliant result may be more valuable than a fast model that ignores an engineering restriction. The benchmark therefore separates final outcome, process efficiency, and constraint compliance.

\subsection{Ground Truth and Reproducibility}

We distinguish deterministic task reproducibility from probabilistic agent reproducibility. The task instance and evaluator are deterministic when the benchmark releases the model files, simulator version, integration settings, random seeds where applicable, fault definitions, semantic-memory content, allowed actions, and scoring code. For a fixed submitted trajectory, the simulator should reproduce the same pass/fail labels, severity scores, voltage extrema, and constraint checks.

The agent layer is probabilistic: LLM sampling, hosted-model updates, tool-call timing, and simulator-recovery choices can change the action sequence even when the task instance is fixed. We therefore define an agent result as reproducible when repeated independent runs under the released harness produce statistically stable values of the declared metrics, rather than identical transcripts. For DMQ, these metrics include full-suite pass rate, waveform-fit measures, number of iterations, and constraint violations. For DSR, they include top-$k$ recall or ranking distance, absolute severity-score and voltage-extrema errors, mitigation success, wall-clock time, and constraint violations. The case studies use ten independent runs and report success rates and median values to make this probabilistic reproducibility explicit.

A practical challenge is that many industry dynamic models are confidential. The benchmark can address this by releasing synthetic and modified public test systems, plus anonymized model packages where possible. For IBR-focused model review, benchmark cases can use generic renewable energy models and injected parameter inconsistencies. For transient stability screening, public systems such as the IEEE 9-bus, IEEE 39-bus, and larger synthetic grids can provide initial testbeds, while the benchmark interface remains extensible to proprietary systems.

\subsection{Limitations and Future Work}

PowerAgentBench-Dyn is a first-step benchmark design rather than a complete replacement for human engineering judgement. The proposed tasks simplify many issues encountered in production studies, including data confidentiality, protection model fidelity, EMT/RMS model consistency, and operator-specific criteria. Future work will expand the benchmark to include more dynamic model families, EMT cases, oscillation-source identification, event replay from PMU data,  and integration with open-source agent frameworks.

Another important direction is human-in-the-loop evaluation. Many real workflows require an engineer to approve a proposed model change or mitigation action. The benchmark can include approval checkpoints where the agent must explain evidence and uncertainty before taking a high-impact action.

\section{Conclusion}

This paper introduces PowerAgentBench-Dyn, a benchmark framework for evaluating agentic AI in power system dynamic studies. The proposed benchmark focuses on tasks that are difficult to solve through deterministic scripting or single-shot optimization, but that are common in real engineering practice. We define two initial benchmark tasks: Dynamic Model Quality Review and Dynamic Security Risk Screening. The first evaluates whether an agent can run model quality tests, diagnose model issues, and recommend constrained corrections. The second evaluates whether an agent can use semantic memory and limited simulations to rank the most risky short-circuit contingencies from an unseen fault dataset, with an extension toward mitigation selection. Together, these tasks provide a foundation for reproducible evaluation of AI agents that interact with dynamic simulation tools and support future power system engineering workflows.

\section*{Disclaimer}
{The views expressed in this paper are the opinion of the
authors and do not reflect the views of PJM Interconnection,
L.L.C. or its Board of Managers, of which Le Xie is a member.}

\bibliographystyle{IEEEtran}
\bibliography{references}

\end{document}